\documentstyle[preprint,aps]{revtex}
\newcommand{\bra}{\langle}
\newcommand{\ket}{\rangle}

\newcommand{\mubar}{\overline{\mu}}
\newcommand{\be}{\begin{equation}}
\newcommand{\ee}{\end{equation}}
\begin{document}
\title{
Spin-Rotation Coupling in Ferromagnetic Clusters
}
\author{
G.F. Bertsch and V.Visuthikraisee
}

\address{
Dept. of Physics and Institute for Nuclear Theory\\
University of Washington\\
Seattle, WA 98195
}

\maketitle

\begin{abstract}
We examine the magnetic response of free clusters considering
the spin direction and the cluster orientation as the
only active degrees of freedom.  The average magnetization
in small fields approaches the Langevin value for paramagnets,
depending on the degree to which the Hamiltonian preserves
symmetries.  Superparamagnetic behavior is not achievable
within models considering only these degrees of freedom. 
\end{abstract}
\section{Introduction}

Experimental studies of magnetism of free atomic clusters
have shown a number of interesting phenomena 
\cite{deheer90,bloomfield91,bloomfield92,deheer93,bloomfield93,
Bi93,bloomfield95}.
Susceptibilities measured in
Stern-Gerlach experiments can be large, with an irregular dependence on the
number of atoms in the cluster.  High susceptibilities are found not only in
normally ferromagnetic elements but also in an element, rhodium
\cite{Co93}, which
is not ferromagnetic in the bulk.  The character of the
susceptibility is also variable. In many cases it is described
phenomenologically as superparamagnetism, which means that besides a
large susceptibility the clusters behave uniformly, independent of
initial state within the thermal ensemble.  The clusters may also
respond with a large variance in their magnetization, with some
clusters even having magnetic moments opposite the field.  Of course,
the limiting
behavior of this kind is the magnetization distribution of
an isolated atom, proportional to the spin projection along the
magnetic field.

     The general theoretical problem we address is whether the
observed behavior of free magnetic clusters can be understood in the
framework of Hamiltonians that treat only rotational and spin
degrees of freedom, neglecting vibrations.  Vibrational coupling is
of course important for the response of warm clusters\cite{deheer93}, but
the question of whether cold clusters can show superparamagnetism is
still open. This paper is an extension 
of earlier work which considered a very simplified Hamiltonian with
only rotational degrees of freedom active \cite{Be94,Be95}.  Large
average susceptibilities were found to be possible, but not
superparamagnetism.

   We shall calculate the cluster magnetism in a quantum mechanical
representation, although the angular momenta
are high enough that classical methods should suffice.  The rotor
is represented in the usual basis $(JKM)$, with $M$ the component of
angular momentum about the $z$  axis and $K$ the component about the 
intrinsic third axis.  The rotor is coupled to spin states in the
basis $(SS_z)$.  Taking the
field direction along the $z$ axis, the component of angular
momentum about that axis, $M+S_z$, is conserved.  In the earlier
studies, the simplified Hamiltonian conserved $K$ as well.

The theoretical magnetic response associated with a particular
Hamiltonian is most conveniently discussed by comparing with
Langevin model of paramagnetism.  Here the spin is
coupled to a nonspecific heat bath to give the following formula
for the magnetization in an external field,
\begin{eqnarray}
   \left\bra\mu_z\right\ket &=& \frac{{\int}_{-\mu_0}^{\mu_0} d\mu_z \ 
                                      \mu_z\exp(\mu_z B/kT)}
				     {{\int}_{-\mu_0}^{\mu_0} d\mu_z 
                                      \exp(\mu_z B/kT)} \\ 
      &=& \mu_0 \Big(  \coth(x) - \frac{1}{x} \Big) 
\label{langevin}  
\end{eqnarray}
Here $\mu_0$ is the intrinsic magnetic moment and the parameter $x$
is the ratio of magnetic energy to thermal energy,  $x=\mu_0 B /k T$. 
This formula is sometimes used to deduce intrinsic magnetic
moments from the observed deflections in the Stern-Gerlach
apparatus. At low field strength the Langevin formula has the limiting
behavior,
\be
  \frac{\bra\mu_z\ket}{\mu_0} = {x \over 3 }.
\ee
The condition $x\ll 1$ is in fact satisfied for the conditions of the cited
experiments, which involved clusters of a few
tens or hundreds of atoms.

Given the Hamiltonian, the magnetic response is calculated as follows.
First, the Hamiltonian is diagonalized for a range of magnetic field
strengths to produce a spectrum of eigenenergies $E_i(B)$.  The magnetic
moments of the states are then calculated using the formula
\be
    \mu_i(B) = \langle iB|\mu_z|iB\rangle = -\frac{dE_i}{dB}.
\label{mu}
\ee
Since eq. (\ref{mu}) does not require the eigenfunctions, a substantial
savings of computation time is possible with respect to a full
diagonalization.  The distribution of eigenvalues is given by the 
expression
\be
P(\mu) = {\sum_i^N P_i \delta(\mu_i -\mu)\over \sum_i^N P_i}.
\label{prob}
\ee
Here $N$ is the dimensionality of the Hamiltonian and $P_i$ are the
probabilities of the states.  We take the adiabatic limit by
assigning the probabilities as the Boltzmann factors of the levels 
in the
absence of a magnetic field,
\be
P_i = \exp(-E_i(0)/kT).
\ee
In applying this, the levels are always ordered and indexed by energy.
Thus the lowest level in zero field sets the Boltzmann factor for
the lowest level at other field strengths.  This assumes that all
level crossings are avoided crossings.  It does appear for the
Hamiltonians we study that the only true level crossings are those
imposed by symmetry ($M$ and possible $K$).  We do not address the
question of whether the field changes take place slowly enough
for the adiabatic assumption to be valid at the avoided level
crossings.

\section{Hamiltonian}

We consider a Hamiltonian model in which the spin orientation is coupled 
to an 
intrinsic axis of the cluster and to an applied field $B$. The Hamiltonian 
can be 
written as
\begin{equation}
	H = \sum_i \frac{J_i^2}{2I_i} - {\lambda\over S^2-S/2} 
\sqrt{4\pi\over 5}\sum_{m} Y_{2m}^{\dagger}
s_{2m} - s_z \mu_0 B 
\label{Ham}
\end{equation} 
where $I_i$'s are the moments of inertia.
The second term is the anisotropy energy, the coupling of the spin to the 
intrinsic axis of the
cluster.  We assume a uniaxial anisotropy having the form of an $L=2$ 
tensor, which would apply to clusters of low symmetry.  A Hamiltonain of
this form was also considered in ref. \cite{Ba91}. The usual
anisotropy energy applying to cubic crystals has the form of
an $L=4$ tensor.

The dimensionalities of the state vector spaces for cluster spin-rotor
wave functions require us to make additional simplifications.  The angular
momenta present in these clusters are of the order of 1000 for a cluster
of 100 iron atoms at room temperature.  It is impractical from a 
numerical standpoint to deal with 
ensembles having such large angular momenta.  Since
the quantization of angular momentum should not play a role at large values,
we believe it is acceptable to use a smaller basis, scaling the parameters
of the Hamiltonian and the temperature appropriately.  Even with a smaller 
basis, the full
Hamiltonian still gives large matrices, so we consider as well a limiting
case where the spin is not an independent dynamical variable, but is
fixed to the cluster axis, the ``locked-spin" model.

We further assume in the locked-spin model that the cluster has two equal
moments of inertia, which we take as the 1- and 2- axes, $I_1=I_2$.
Then in the usual $|JKM\rangle$ basis $K$ is conserved by the rotational
Hamiltonian.  Its nonvanishing matrix elements are the diagonal elements,
given by
\be
\langle JKM| \sum_i \frac{J_i^2}{2I_i}|JKM\rangle = {J(J+1)\over 2 I_1} +
{(I_1-I_3)K^2\over 2 I_1I_3}
\ee
A fully asymmetric rotor requires an additional term in the Hamiltonian
with matrix elements proportional to 
\be
\label{asymm}
\langle J\,K+2\,M| J^2_1-J^2_2 |JKM\rangle =
\sqrt{(J(J+1)-K(K+1))(J(J+1)-(K+1)(K+2))}
\ee
We will briefly consider this term later.

The rotor is coupled to the magnetic field through the spin, whose
orientation with respect to the intrinsic 3-axis is denoted by $\theta$.
The matrix elements of the magnetic field term in the Hamiltonian are
diagonal in $M$ but not necessarily in $J$ or $K$.  They are given by
\begin{eqnarray}
        \bra JKM|\mu_z B | J'K'M' \ket 
	&=& \delta_{M,M'}\mu_0 B \sqrt{ \frac{2J'+1}{2J+1} } \, 
         \bra J'M10|JM\ket 
	\nonumber \\
        & & \times \!\!\left\{ \cos\!\theta \bra J'K'10|JK\ket 
        + \frac{\sin\!\theta}{\sqrt{2}} 
    \!\left[\: \bra J'K'1-\!\!1|JK\ket\! - \!\bra J'K'11|JK\ket \:\right] 
     \right\} \; .\ \ 
\label{mag1}
\end{eqnarray}

The more  general Hamiltonian allowing intermediate coupling of the spin
is constructed in the $|JKMSS_z\rangle$ basis.  The magnetic energy is
trivial in this basis.  It depends only on $S_z$, with the matrix
element $\langle S_z|\mu_0 B|S_z\rangle = \mu_0 B S_z$.  
The anisotropy energy is given by
\begin{equation}
\begin{array}{l}
    \bra JKMSS_Z | \sum_{m} Y_{2m}^{\dagger}
s_{2m} |J'K'M'SS_z' \ket
 = (-1)^{2S+S'_z-S_z} \sqrt{5\over 4 \pi}\sqrt{(2J+1)(S+1)(2S+3)S(2S-1)\over
4(2J'+1)}     \\ \\
   \,\,\,\,\,\,\,\,\,
     \delta_{K,K'}\delta_{M+S_z,M'+S_z'}
   \bra J\,\, M\,\, 2\,\,\, (M'\! -\! M) | J'\,\, M'\ket 
   \bra J\,\, K\,\, 2\,\, 0 | J'\,\, K \ket  
    \bra S\,\, S_z\,\, 2\,\,\, (M\! -\! M') | S\,\, S'_z \ket 
\end{array}
\label{mat;ani}
\end{equation}

\section{Results:  Locked Spin model}

The locked-spin model for a spherical rotor was treated in ref.
\cite{bloomfield91,Be94,Be95,Ma93}.
In this model, $K$ is a conserved quantum number if the intrinsic 3-axis
is taken along the spin direction. 
The previous work showed that the adiabatic ensemble produces an average 
susceptibility in low fields of
2/3 of the Langevin value.  The distribution $P(\mu)$ is very
broad, however, far from the idealized superparamagnet.  
A weak point of the model is its assumed $K$ conservation: one could 
argue $K$ conservation is too restrictive to
allow adequate mixing of the spin with the other degrees of freedom.  We
shall find that breaking $K$ does in fact have a significant effect on the
average magnetization, but not on the spread.  

We first remark that changing the
Hamiltonian by making the rotor deformed has negligible effect on the
distribution, in the absence of $K$ mixing.  When $K$ is conserved, the 
deformed rotor has the
same wave function as the spherical rotor.  The magnetic moments of the
states are thus the same, only the occupation probabilities differ.  The
thermal ensemble would then favor states a somewhat different distribution
of $K$ values, but the overall effect is small.  This is illustrated with
the moment distributions for the 
deformed symmetric rotor shown in Fig. 1.    The curves were calculated
with an ensemble having a cutoff angular momentum of 40 for the rotor.  
The temperature
was fixed at $kT=100$ in units of $I_1^{-1}$ and the magnetic field strength
has the same value, $\mu_0B=100I_1^{-1}$. To display the distribution, eq.
(\ref{prob}) was smeared with
a Gaussian of the form $\exp(-(\mu-\mu_i)^2/(\Delta \mu)^2)$ with
$ \Delta \mu = 0.05\mu_0$. 
The different curves show results for Hamiltonians with moment of inertia 
ratios ranging from
1/2 to 2. One sees that there is hardly any change in the distribution of
magnetic moments.  The average magnetization changes by less than 10\%
over this extreme range of deformations.   

Without discussing in detail the $K$-mixing in the Hamiltonian,
it is possible to see in a crude model how it affects the magnetization.
We compute the energy with a Hamiltonian that conserves $K$, but 
combine the ensembles for different $K$ (but the same $M$) into a
common ensemble in eq. (\ref{prob}).  This assumes that the K-mixing makes all level 
crossings avoided crossings, but that it is not strong enough to
affect the states between crossing points.  This ensemble produces
a higher average magnetization, as might be expected:  the crossing
of states of different $K$ allows the probability to jump to states
of larger magnetic moment.  
In Fig. 2 we show the average magnetization of this mixed-$K$ ensemble
compared with the Langevin and with the fixed-$K$ locked-spin model.  One
sees that the mixed-$K$ ensemble has a susceptibility close to the 
Langevin model for field strengths below $x=0.2$.  However, the
spread in moments is very similar to the fixed-$K$ ensemble. 

We now discuss the behavior of eigenstates in Hamiltonians with
mixed $K$.  In the locked-spin models, the $K$ mixing can be 
induced by a tipped spin with $\theta$ nonzero, or by a fully
asymmetric rotor with the additional term eq. (\ref{asymm}) in the 
Hamiltonian.
The effect of the tipped spin may be seen in the energy level
plot in Fig. 3.  The plot show the energies of the $M=1$ states with spin
angle tipped by $5^\circ$ with respect to the symmetry axis. Several
avoided level crossings may be seen.
For example, the uppermost crossing involves a
transition between states that have quantum numbers $(JK) = (20)$
and $(JK) = (1,-1)$ in zero field.  We can see that the
effect of the $K$-mixing is limited to regions very close
to the level crossings; the spectrum is hardly changed otherwise.
This means that the magnetic moments of the levels are unaffected
except for special values of the field.  The consequence is
that the dispersion in magnetic moments is very similar to
that of the fixed $K$ ensemble.  This may be seen in 
Fig. 4, where we show the magnetization distribution for
a range of tipping angles.  There is very little difference between
the different Hamiltonians. The average magnetization from the diagonalized
Hamiltonians with mixed $K$ is also very close to the mixed-$K$
ensemble shown in Fig. 2.

Returning to the spread in magnetization distribution, it appears that the
locked-spin Hamiltonian cannot explain superparamagnetic behavior.  The
one caveat in this statement is that we have not considered fully 
asymmetric Hamiltonians.  The perturbation eq. (9) does not produce a
large mixing of the wave functions, and seems unlikely to have a
significant effect on the widths.

\section{Results: intermediate coupling}

We now generalize from the locked-spin Hamiltonians to consider
the spin as an independent degree of freedom.  We shall treat
the intermediate coupling spin-rotor Hamlitonian assuming that
the inertia tensor is axially symmetric.  As we saw in the
last section, the breaking of $K$ symmetry has no qualitative
effect on the magnetic response, but only changes somewhat the
magnitude of the susceptibility coefficient.  We study the ensemble 
with a maximum orbital angular momentum 
of $J_{max}=30$ and the spin value $S=1$.  Despite the modest size
of these numbers, this space requires diagonalization of about
4000 matrices of dimensionality ranging up to about 100.  The
ensembles are generated with a temperature of $T=100/I_1$ as 
before.

The first task is to see whether the locked-spin results can be
reproduced with a spin $S$ as small as $S=1$. For this purpose
we calculate the moment distribution taking a large value for
the anisotropy coupling, $\lambda= 5 \mu B$.   Typical results are
shown in Fig. 5.  In plotting the distribution, the magnetic moments 
have been scaled so
that the maximum is $\pm1$ in either model.  We see that the
two curves are practically identical, giving us confidence in
using the Hamiltonian with a small value for $S$.  The moment 
distribution in a weak coupling situation,
$\lambda = 0.01 \mu_0 B$, is shown in Fig. 6.  Here the spin 
behaves as though it were uncoupled, so one sees the three $M$-states 
as in the atomic deflection.  Notice however that the occupation
probabilities are unequal. The probabilities are shifted at the level
crossings, which become avoided crossings due to the
(small) anisotropy term. At intermediate values of the coupling, the 
distribution changes
smoothly from the three-peak structure in Fig. 6 to the typical
locked-spin shape of Fig. 5.  In all cases the distribution 
remains broad.

The field dependence of the magnetization shows a more complex behavior 
with intermediate
coupling than we found for the locked spin.   There are
several distinct regions of behavior, depending on the relative sizes
of the magnetic energy with respect to the anisotropy energy, and the 
rotational
frequency with respect to the Larmor frequency.    In extremely low
fields, with magnetic energies less than the anisotropy energy, the
spin should behave in a similar way to the locked-spin model.
When the magnetic field energy becomes
larger than the anisotropy energy, the spin decouples from the
rotational motion and we should see the individual $S_z$ states as in Fig. 6.
These two regimes are illustrated in Fig. 7, showing the magnetization
distribution for $x=0.005$ and 0.02 for a model with weak coupling,
$\lambda=0.01 kT$.  The distribution in the
lowest field, shown by the solid line, is broad with a peak
at $\mu=0$.  This is similar to the locked-spin behavior, $P(\mu)\sim
\log(\mu/\mu_0)$. At $x=0.02$, where the magnetic field is twice
the anisotropy energy, the distribution has already dissolved into
the 3-peak structure of the decoupled $S=1$ states.  The
behavior of the average magnetization as a function of $x$ is shown
in Fig. 8. In the extreme low-field limit, $\mu_0 B  <
\lambda $, the magnetization is close to the weak-field Langevin
function,  eq. (3).  The spin decouples from the rotational motion when
$B$ exceeds $\lambda$, as already mentioned. From Fig. 8 it
may be seen that the average magnetization increases rather slowly
between $\mu_0 B \sim 0.01 kT$ and $0.08 kT$.  Above that point the
magnetization rises to even exceed the Langevin value.  We believe
that this change of behavior is associated with level crossings
of the type $(J\,S_z) \rightarrow (J\pm1\,\,S_z\mp1)$, corresponding
to the matching of rotational and Larmor frequencies.  A change
of behavior at this matching point has been suggested previously
\cite{Bi93,Be93}.  

It is interesting to examine the susceptibility in the third region
in more detail.  Here the magnetization is still far below the saturation
condition $\mu_0 B \sim kT$, and it is roughly linear in the field.
We make a linear fit to the magnetization with the function,
\be
\mu = a  \mu_0 x,
\ee 
The dependence of the fit coefficient $a$ on magnetic field is plotted in 
Fig. 9. We see that the behavior like the the locked-spin model only if 
the anisotropy energy is substantially above $kT$.  Below that, the 
coefficient exceeds
the Langevin value, $a=1/3$, as we saw on Fig. 7.    

Our findings for the magnetic response are quite different from those
of ref. \cite{Ba91}, where a similar Hamiltonian was considered.  In
that work, the magnetization was found to be independent of the
temperature but to depend strongly on the anisotropy energy.  However,
these authors did not treat the adiabatic ensemble, but rather a
mixed ensemble in which the magnetic field was turned on suddenly.

We conclude this section with an estimate of the magnitude of 
$\lambda$ for iron clusters.  The cubic anisotropy constant $K$
for bulk iron is measured as \cite{La} $K\approx4 \times 10^5 {\rm~ergs/cm}^3
= 5 \times 10^{-4}$ eV/atom.  For a cluster of 20 iron atoms this
is $2.5 \times 10^{-4}$ eV, which scaled to room temperature 
implies  $\lambda\sim 0.01 kT$.  Thus the weak coupling limit
we discussed above should apply to moderate sized iron clusters
at room temperature.  A typical rotational frequency is given
by $\langle J^2\rangle^{1/2}/I \sim (kT/I)^{1/2}$.  This has
a magnitude of about $ 10^{-4}$ eV/$\hbar$ for a cluster of 
20 iron atoms, which is the order of magnitude of the Larmor
frequency in a 1 T field.  Thus, the model predicts that light
mass iron clusters would behave with the spin nearly decoupled
from the rotational motion and with a fairly weak average
susceptibility.  This is certainly contrary to experiment, which
has never demonstrated atomic-like behavior in clusters larger
than diatomic.

\section{Conclusion}

Our study of the magnetic response of rotors coupled to spin is
guided by the question, to what extent can a thermal ensemble of
rotors play the role of the thermal bath that one assumes in the
Langevin model?   One's intuition may be that the rotor has
far too few degrees of freedom to serve as a heat bath.  Nevertheless,
the low-field magnetic susceptibility of the thermal bath is
reproduced on average by the deformed rotor with a locked and
tipped spin axis.  It is also obtained for intermediate coupling
of spin to the rotor, except in a small interval of field strengths.

However, superparamagnetism implies not only that the average
moment is given by the Langevin formula, but all clusters should
behave identically.  Here the adiabatic rotor ensemble behaves
quite differently, always having a large variance in the magnetization.

Let us ask now, what would be needed in the spin-rotor Hamiltonian to 
achieve superparamagnetism?  Clearly there would have to be
much stronger $K$ mixing than can be generated by the spin-rotor
Hamiltonian.   Recall in Fig. 3, the energy
levels have different slopes with large stretches of constant
moment separated by small intervals of the level crossings.  

In the current idiom, we may speak of this as a weakly perturbed
regular Hamiltonian.  To achieve superparamagnetism, the 
Hamiltonian would have to be chaotic.  Quantum mechanical
chaos is a regularity and evenness in the energy spectra associated with
the complete mixing of wave functions.  In such a spectrum, the
slopes of the levels would be similar and governed by the
average behavior of the levels in that region of energy.  
When
the Hamiltonian is chaotic, the slopes have universal correlation
properties \cite{Al93,Ku96}.  This behavior would require
additional terms in the Hamiltonian eq.(\ref{Ham}).  A minimum 
condition for chaotic behavior is that the off-diagonal matrix
elements exceed the level spacing of the regular part of the
Hamiltonian \cite{Zi83}.

It is expected that the coupling of the spin to vibrations could
provide the thermal bath that would produce
superparamagnetism\cite{Kh91,Be94}.  However,
it has not been demonstrated that this mechanism actually would
work in finite-size clusters.  The first issue is that 
vibrational transitions would have to be at the same energy as
the spin flip in the external field.  This is to allow off-diagonal
matrix elements to nearby levels. The maximum vibrational
frequency has the order of magnitude of the Debye energy
$\hbar\omega_D$; the spacing of vibrational levels $\Delta E_{vib}
\sim \hbar\omega_D/3N$, where $N$ is the number of atoms in the
cluster. In a 1 T field, the spin flip energy is $10^{-4}$ eV.  
With a Debye energy in iron of $6\times 10^{-3}$ eV, the vibrational
level spacing is smaller than the spin-flip energy for clusters
larger than $N=20$.  Spin relaxation in the bulk takes
place mainly through Raman scattering of phonons. In this case
one is interested in the difference of two vibrational energies,
and the vibrational level spacing again sets the scale.  The next condition
is that the average off-diagonal matrix element of the vibrational coupling
should exceed the level spacing.  This will always be satisfied for
sufficiently large clusters, because $V$ and $\Delta E$ vary 
as $V\sim \sqrt{(\Delta E)}$ as the size of system varies. This
follows from the requirement that the relaxation
time is independent of the size of the system for large $N$.
Unfortunately, the coupling is not well enough known at present to make a
numerical estimate.

\section{acknowledgment}
We acknowledge discussions with N. Onishi and A. Bulgac.
This work was supported by the Dept. of Energy under Grant 
FG06-90ER40561.

\begin{figure}
\caption{Comparison of magnetization distributions of spherical clusters 
and deformed clusters, for a range of inertia ratios $I_1/I_3$.}
\label{fig3.3}
\end{figure}

\begin{figure}
\caption{Comparison of average magnetization of mixed-$K$ ensemble with
fixed-$K$ ensemble and with the Langevin function, eq. (1).}
\label{fig2}
\end{figure}

\begin{figure}
\caption{Energy levels for locked-spin Hamiltonian as a function of
magnetic field, for low $M=1$ states. The spin is tipped by $5^\circ$ with 
respect to
the symmetry axis.}
\label{fig3}
\end{figure}

\begin{figure}
\caption{Magnetization distribution of deformed clusters, $I_1/I_3 = 2$, 
with three different spin angles $\theta= 0^{\circ}, 45^{\circ}$, and 
$90^{\circ}$.}
\label{fig4}
\end{figure}

\begin{figure}
\caption{Magnetization distribution of clusters in the anisotropic model, 
showing the strong coupling limit. Here the spin ($s=1$) is coupled to the 
easy axes, with $\lambda = 5 \mu B$ and $x=2$. The distribution is close to 
the result of locked-spin model (dashed line). The distribution is plotted 
as a function of the rescaled magnetization $\mubar$.}
\label{fig5}
\end{figure}

\begin{figure}
\caption{Magnetization distribution for the weakly coupled anisotropic
Hamiltonian.  The spin value is $S=1$, giving three peaks.}
\label{fig6}
\end{figure}
\begin{figure}
\caption{Magnetization distribution for weakly coupled spin to rotor,
$\lambda=0.01kT$.  The solid curve shows the distribution in a very
weak field, $x=0.005$, and the dashed line shows the distribution
in a field $x=0$, somewhat stronger than the anisotropy energy. }
\label{fig7}
\end{figure}
\begin{figure}
\caption{Average magnetization of the anisotropic model for $\lambda=0.01kT$, 
as a function of Langevin parameter $x=\mu B/kT$.}
\label{fig8}
\end{figure}
\begin{figure}
\caption{Fit of the low-field adiabatic susceptibility coefficient
$a$ in eq. (12) as a function of the anisotropy energy.  The temperature is
fixed at $T=50$.}
\label{fig9}
\end{figure}

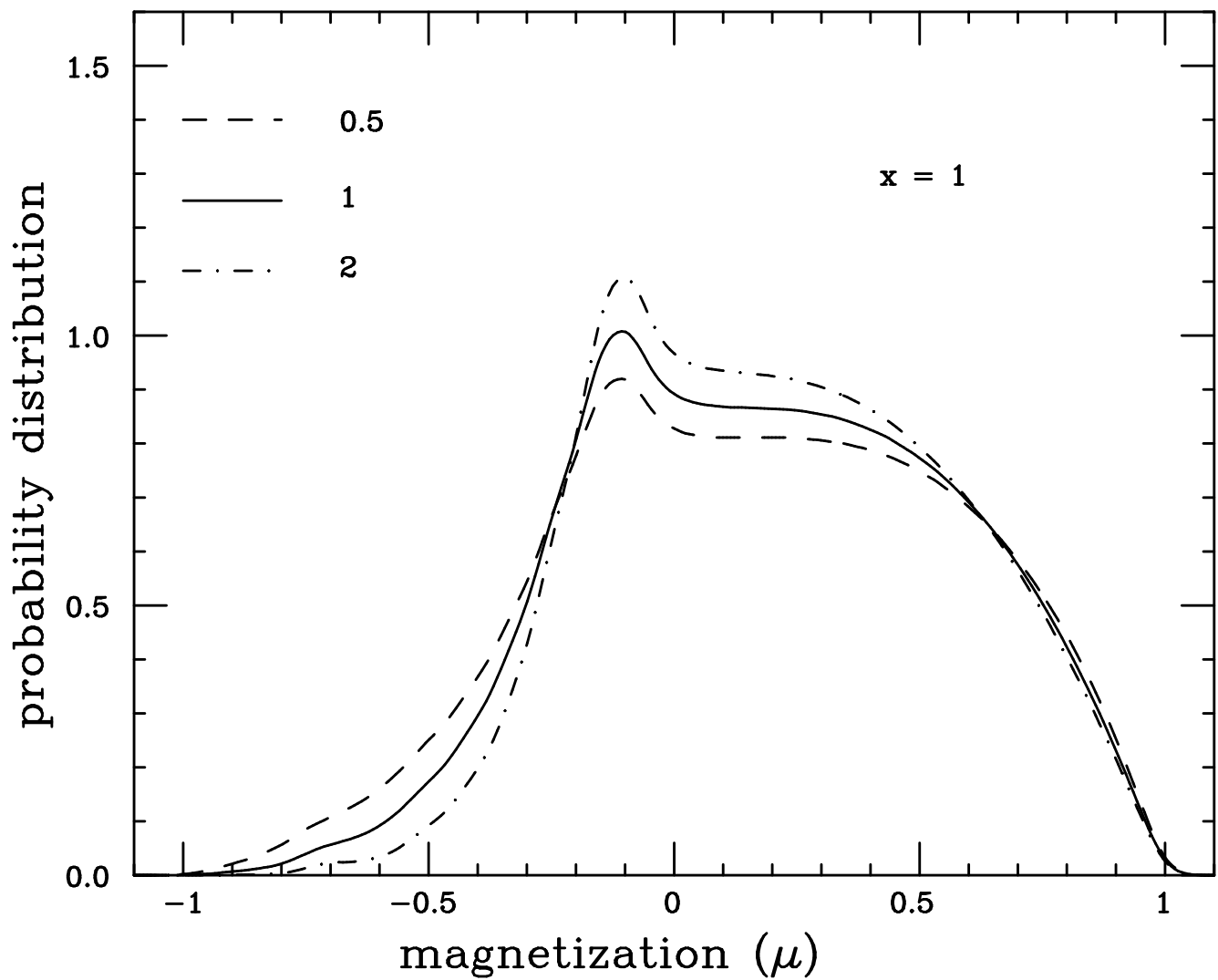
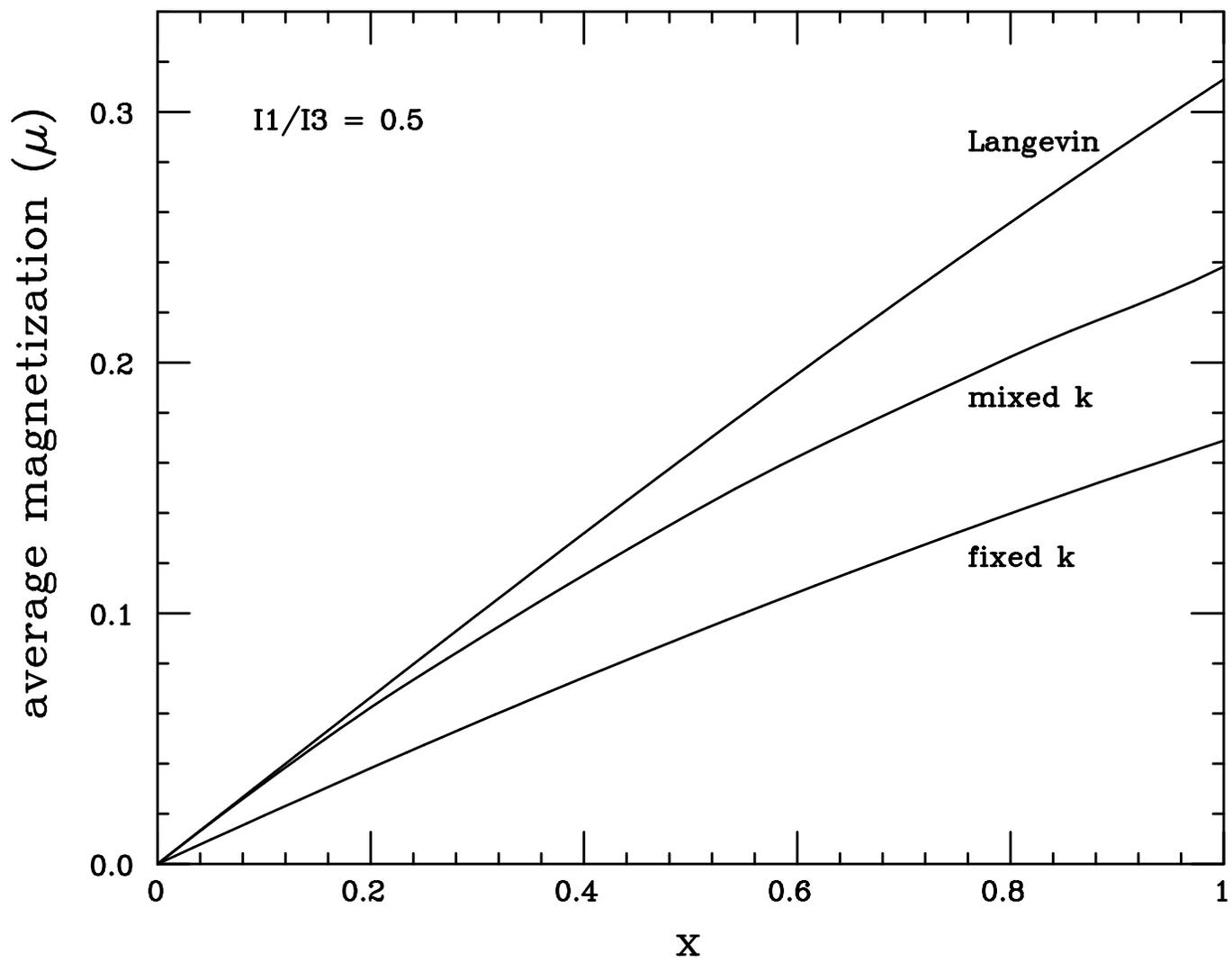
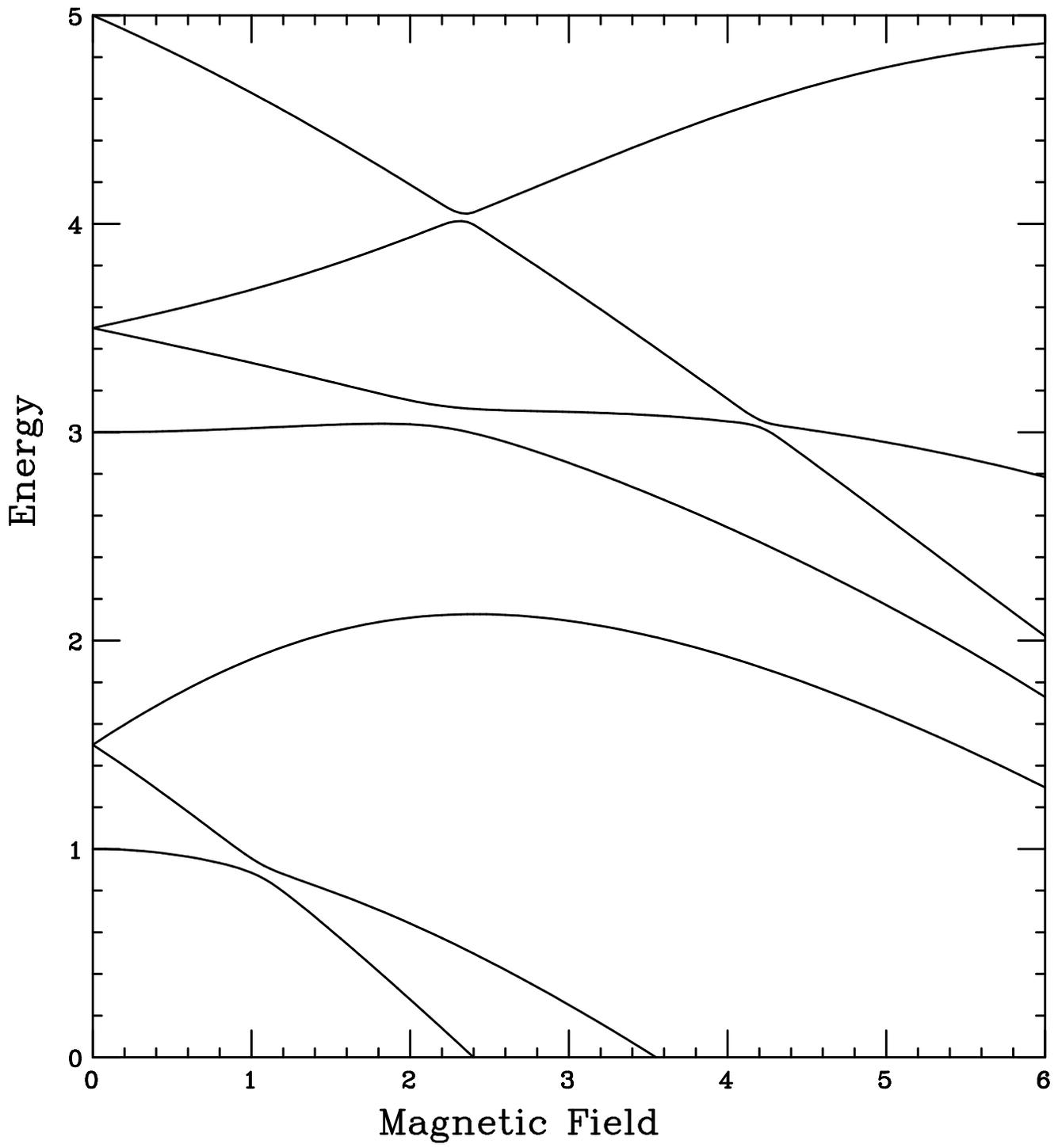
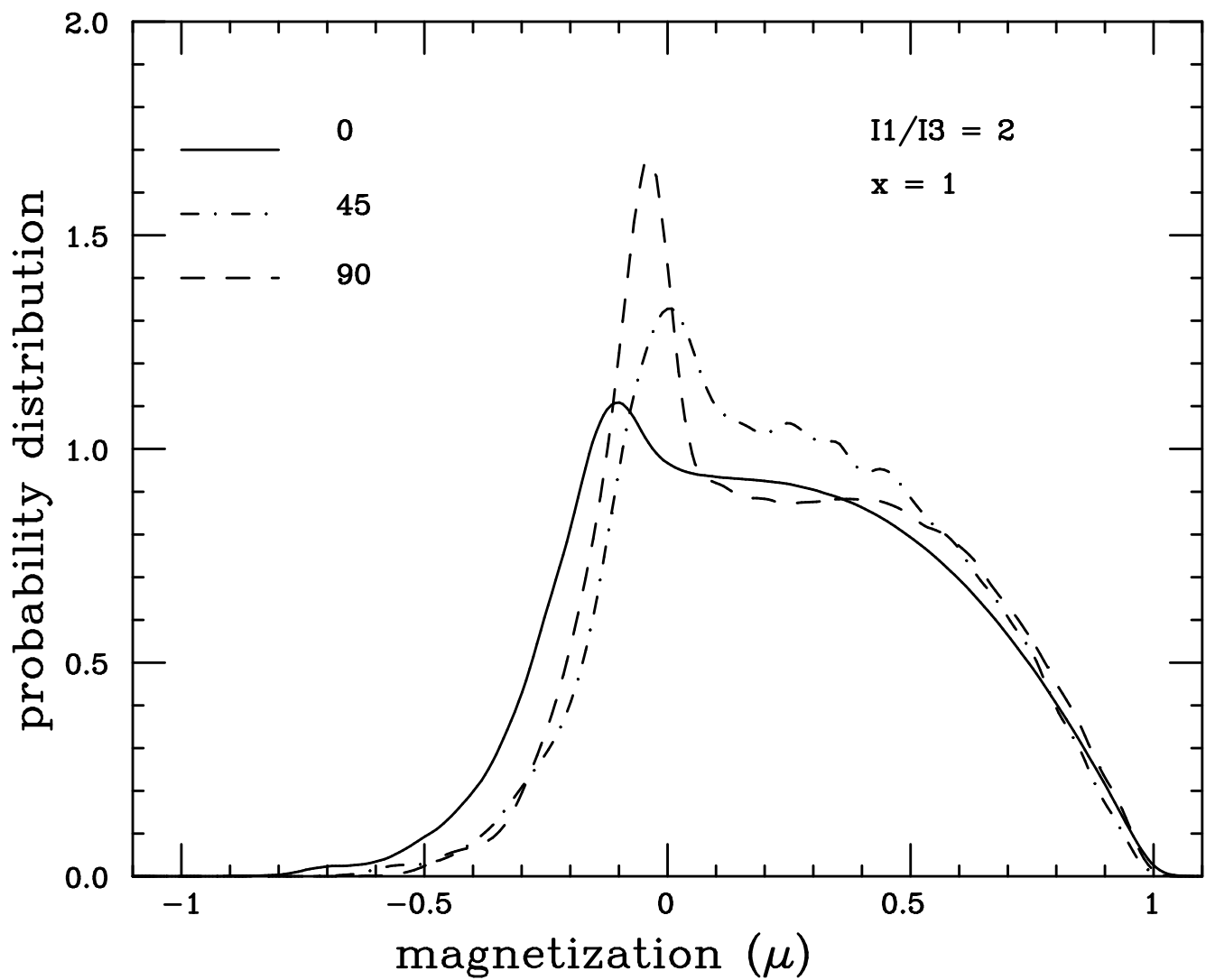
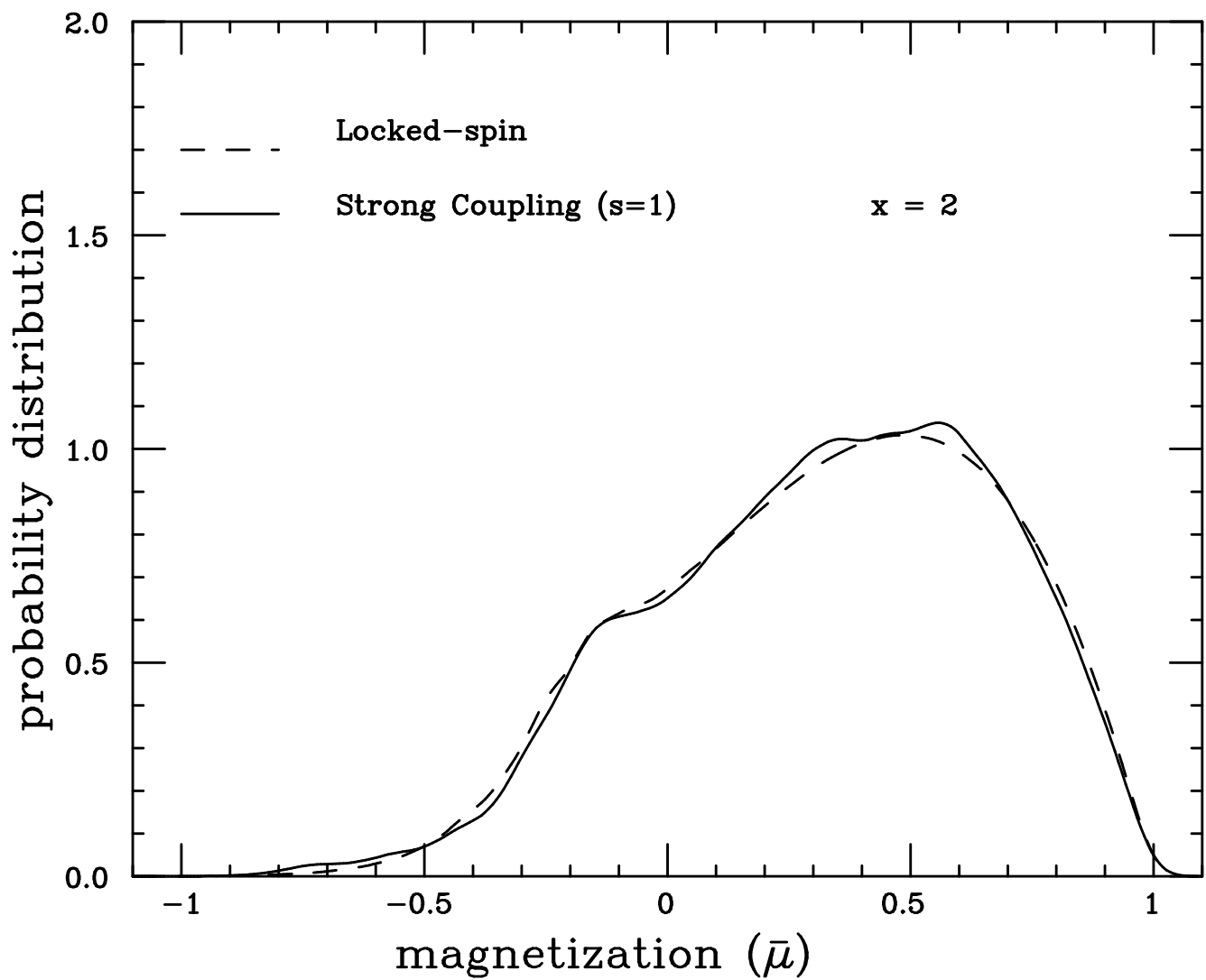
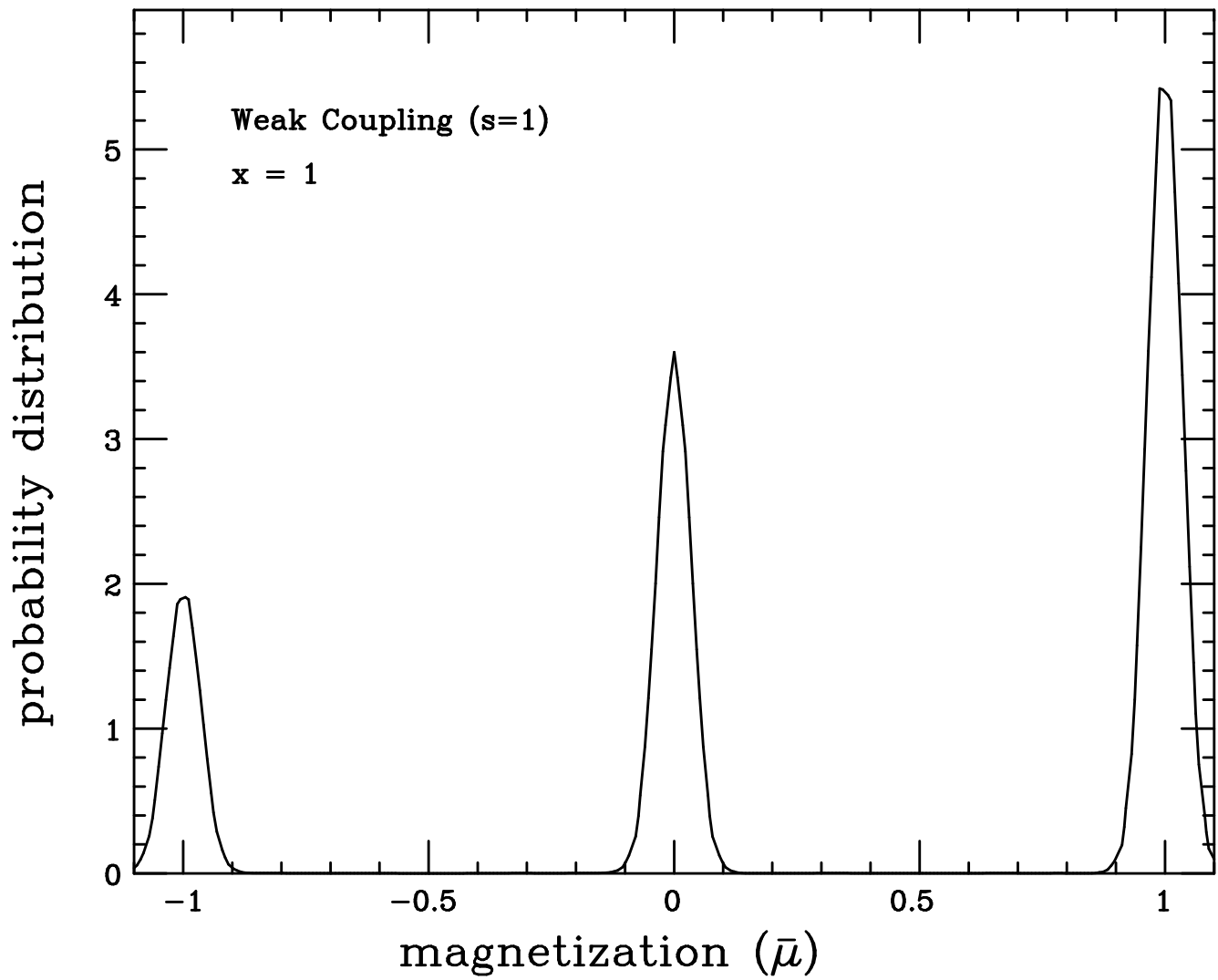
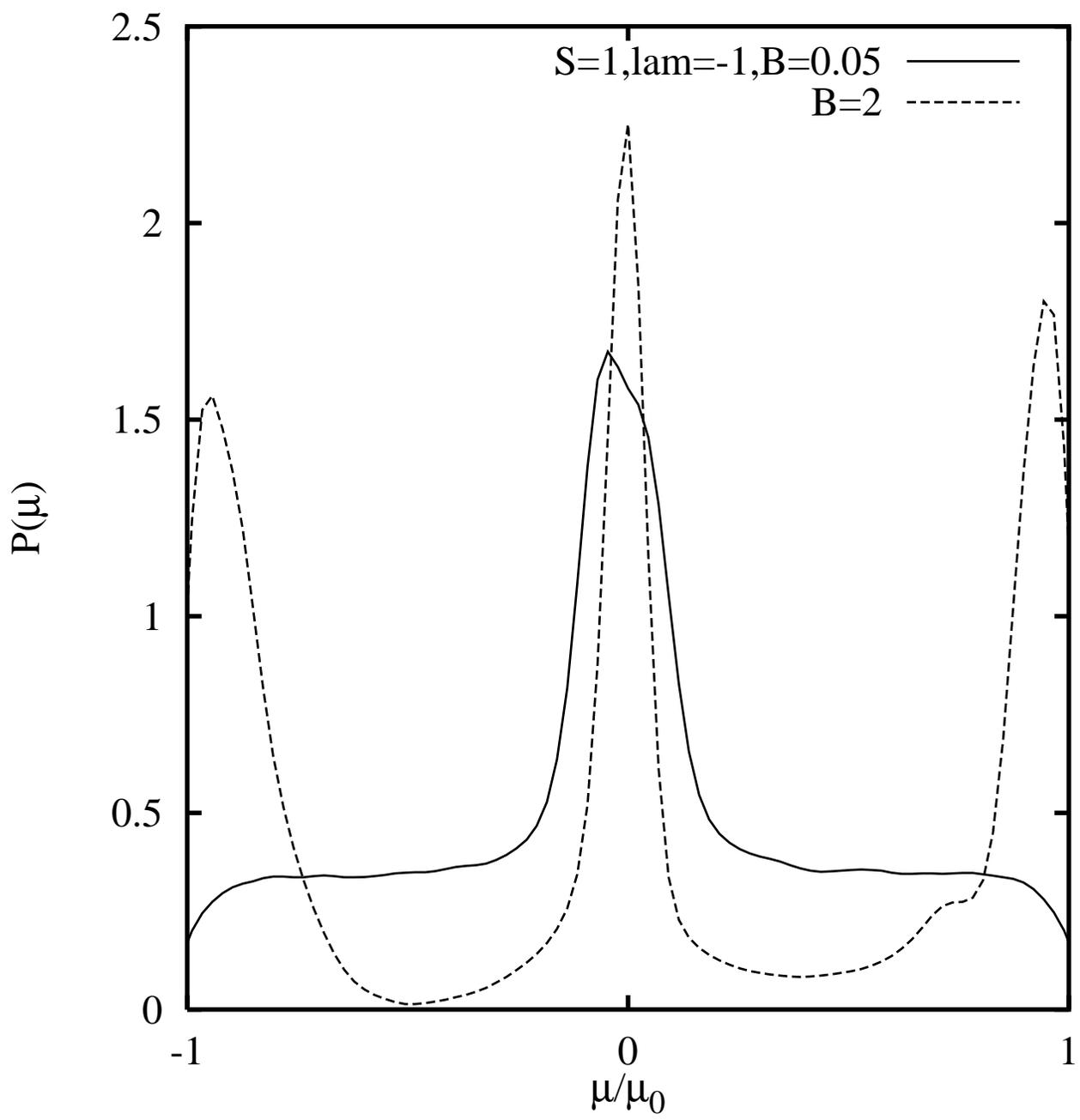
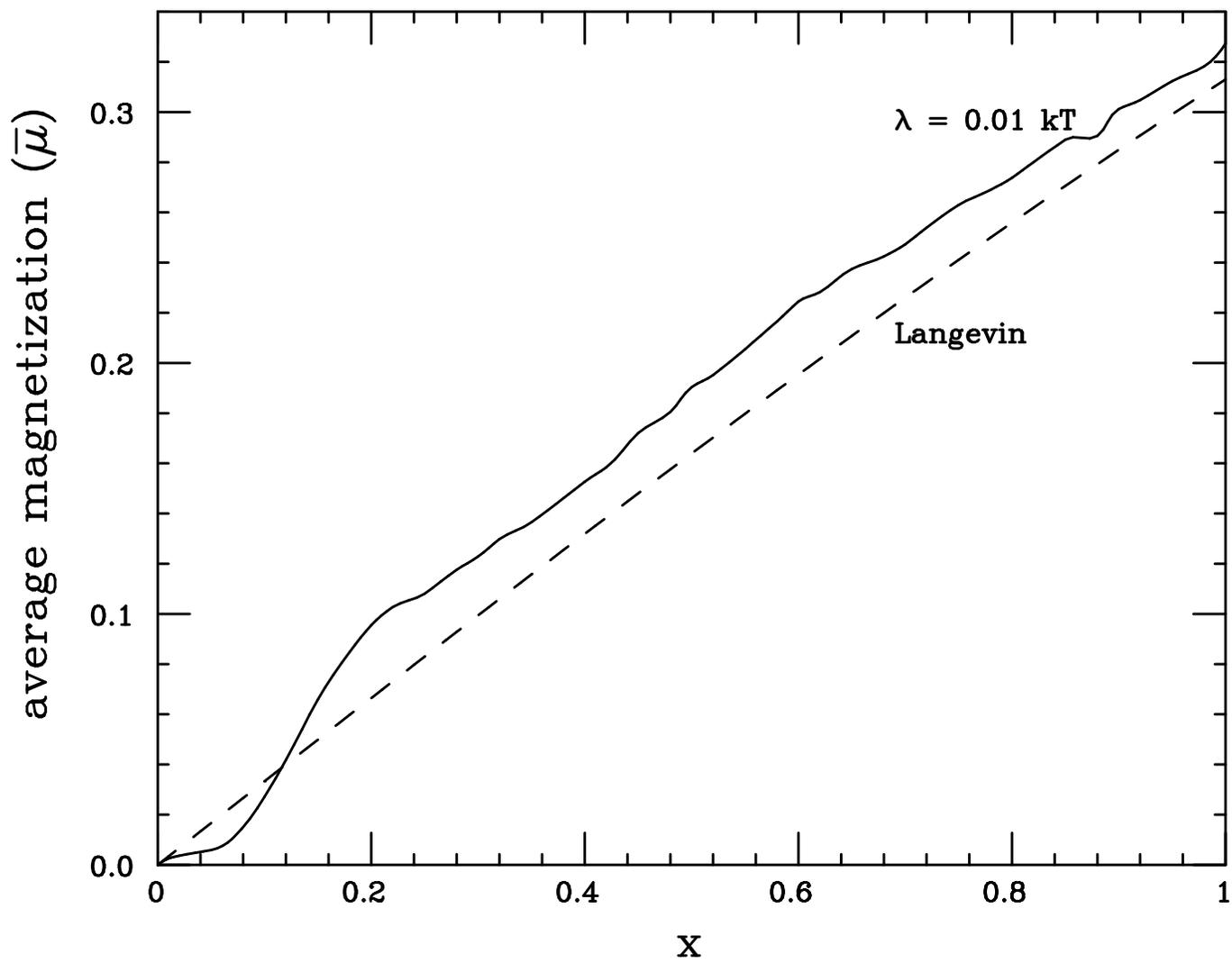
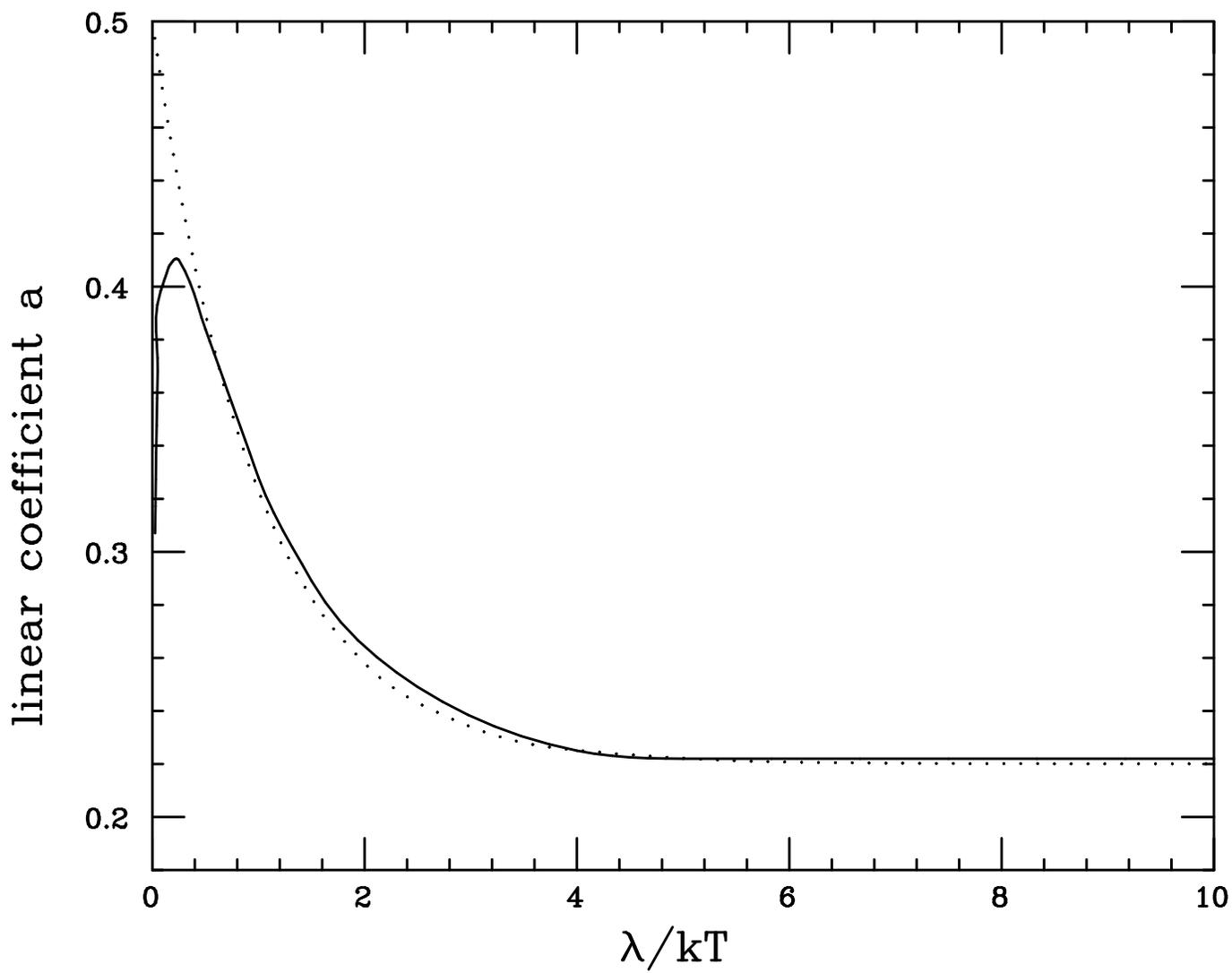
\end{document}